\documentclass{article}
\usepackage{times}
\usepackage{amsfonts}
\usepackage{graphicx}
\usepackage[pdfmark]{hyperref}
\begin{document}
\noindent
{\Large   NONCOMMUTATIVE SPACE  AND THE LOW--ENERGY PHYSICS  OF QUASICRYSTALS}
\vskip1cm
\noindent
{\bf L. Monreal${}^{a}$, P. Fern\'andez de C\'ordoba${}^{a}$, A. Ferrando${}^{a,b}$ and J. M. Isidro${}^{a}$\footnote{Corresponding author. Email address: {\tt joissan@mat.upv.es}.}}\\
${}^{a}$Interdisciplinary Modeling Group, Intertech \\
Instituto Universitario de Matem\'atica Pura y Aplicada, IUMPA\\
Departamento de Matem\'atica Aplicada, Universidad Polit\'ecnica de Valencia,\\
Camino de Vera s/n, Valencia 46022, Spain\\
and\\
${}^{b}$Departamento de \'Optica, Universidad de Valencia, \\
Dr. Moliner 50, 46100 Burjasot, Spain\\

%\vskip1cm
%\noindent
%\today

\vskip1cm
\noindent
{\bf Abstract}  We prove that the effective low--energy, nonlinear Schroedinger equation for a particle in the presence of a quasiperiodic potential is the potential--free, nonlinear Schroedinger equation on noncommutative space.  Thus quasiperiodicity of the potential can be traded for space noncommutativity when describing the envelope wave of the initial quasiperiodic wave. 
\vskip1cm
\section{Introduction}\label{grund}

Ever since their discovery in 1984 \cite{SHECHTMAN}, quasicrystals have challenged physicists and mathematicians alike \cite{SENECHAL, BAAKE}. Recent approaches to the problem of electron propagation and Bloch theory in quasicrystals have made use of techniques borrowed from noncommutative geometry \cite{CONNES, LANDI, YPMA}. In this paper we will apply such techniques to study the propagation of wave pulses within quasiperiodic media, as described by the nonlinear Schroedinger equation.  Let us first present the elements of noncommutative geometry that will be needed later; a nice introductory review for physicists is \cite{SZABO}.

The noncommutative space $\mathbb{R}^N$ is defined by coordinates $x^{i}$, $i=1,\ldots, N$, satisfying the commutation relation
\begin{equation}
[x^{i},x^j]={\rm i}L^2\theta^{ij},\qquad i,j=1,\ldots, N,
\label{kett}
\end{equation}
where $L$ is a length scale and $\theta^{ij}$ a constant, dimensionless, real antisymmetric tensor. The usual pointwise product of functions $f(x)\cdot g(x)$ on commutative $\mathbb{R}^{N}$ is replaced,  on noncommutative $\mathbb{R}^{N}$, with the (associative, noncommutative) $\star$--product,
\begin{equation}
(f\star g)(x):=f(x)\,\exp\left(\frac{{\rm i}}{2}\overleftarrow\partial_i\theta^{ij}\overrightarrow\partial_j\right)\,g(x).
\label{stern}
\end{equation}
Given the Fourier expansions on commutative $\mathbb{R}^N$ 
\begin{equation}
f({x})=\frac{1}{(2\pi)^{N/2}}\int{\rm d}{k}\,\tilde f({ k}){\rm e}^{-{\rm i}{kx}}, \qquad \tilde f({k})=\frac{1}{(2\pi)^{N/2}}\int{\rm d}{x}\,f(x){\rm e}^{{\rm i}
{kx}},
\label{konvv}
\end{equation}
where $kx:=k^jx^j$, eqns. (\ref{stern}) and (\ref{konvv}) imply that the pointwise multiplication law $\tilde f({k})\cdot\tilde g({q})$ for Fourier modes  on commutative $\mathbb{R}^N$ is replaced, on noncommutative $\mathbb{R}^N$, with
\begin{equation}
\tilde f({k})\tilde g({q})\exp\left(-\frac{{\rm i}}{2}{k}\times{q}\right), \qquad {k}\times{q}:=k_{i}\theta^{ij}q_j.
\label{pleee}
\end{equation}
It follows that
\begin{equation}
\int{\rm d}x(f\star g)(x)=\int{\rm d}xf(x)g(x)
\label{menta}
\end{equation}
for any two functions $f,g$ on $\mathbb{R}^N$ possessing a Fourier transform.

\section{The noncommutative wave equation}\label{anfang}

The nonlinear Schroedinger equation \cite{NLS} is known to provide an effective description of numerous low--energy phenomena, from optics to condensed--matter physics. Let us consider the nonlinear Schroedinger equation for a wavefunction $\psi=\psi(t;x^1,\ldots, x^N)$ on $\mathbb{R}\times\mathbb{R}^N$:
\begin{equation}
{\rm i}\frac{\partial\psi}{\partial t}+\frac{1}{2}\nabla^2\psi+g\vert\psi\vert^2\psi=0.
\label{nlsq}
\end{equation}
This is the equation of motion for the action 
\begin{equation}
S=\int{\rm d}t{\rm d}x\,\left({\rm i}\psi^*\frac{\partial\psi}{\partial t}-\frac{1}{2}\nabla\psi^*\nabla\psi+\frac{g}{2}\vert\psi\vert^4\right).
\label{ncakk}
\end{equation}
In what follows we will assume that the space $\mathbb{R}^N$ becomes noncommutative, while the time $t$ will continue to commute with the space coordinates $x^j$.
On noncommutative $\mathbb{R}^N$ the action (\ref{ncakk}) thus becomes
\begin{equation}
S_{\rm nc}=\int{\rm d}t{\rm d}x\,\left({\rm i}\psi^*\star\frac{\partial\psi}{\partial t}-\frac{1}{2}\nabla\psi^*\star\nabla\psi+\frac{g}{2}\psi^*\star\psi\star\psi^*\star\psi\right),
\label{ncakknon}
\end{equation}
which by eqn. (\ref{menta}) reduces to
\begin{equation}
S_{\rm nc}=\int{\rm d}t{\rm d}x\,\left({\rm i}\psi^*\frac{\partial\psi}{\partial t}-\frac{1}{2}\nabla\psi^*\nabla\psi+\frac{g}{2}\psi^*\star\psi\star\psi^*\star\psi\right).
\label{ncakknonmm}
\end{equation}
Thus only the selfinteraction term is sensitive to the noncommutativity, while the free terms in the action remain as on commutative space. The equation of motion  corresponding to (\ref{ncakknonmm}) is \cite{GREECE}
\begin{equation}
{\rm i}\frac{\partial\psi}{\partial t}+\frac{1}{2}\nabla^2\psi+g\psi^*\star\psi\star\psi=0.
\label{ncmmnlsq}
\end{equation}
To first order in the noncommutativity parameter $\theta$ we have
\begin{equation}
\psi^*\star\psi\star\psi=\vert\psi\vert^2\psi+{\rm i}\left(\nabla\psi^*\times\nabla\psi\right)\psi+O(\theta^2).
\label{trix}
\end{equation}
Hence, in the limit of weak noncommutativity, eqn. (\ref{ncmmnlsq})  reduces to
\begin{equation}
{\rm i}\frac{\partial\psi}{\partial t}+\frac{1}{2}\nabla^2\psi+g\vert\psi\vert^2\psi+{\rm i}g\left(\nabla\psi^*\times\nabla\psi\right)\psi=0.
\label{nlsqzzxz}
\end{equation}
We will refer to (\ref {ncmmnlsq}) as the nonlinear Schroedinger equation on noncommutative $\mathbb{R}^N$, and to (\ref{nlsqzzxz}) as its weakly noncommutative limit.

\section{The effective equation for long--range amplitudes}\label{anspruch}

Let us consider the nonlinear Schroedinger equation on a commutative space and in the presence of a static potential $V(x)$:
\begin{equation}
{\rm i}\frac{\partial\Psi}{\partial t}+\frac{1}{2}\nabla^2\Psi-V(x)\Psi+g\vert\Psi\vert^2\Psi=0.
\label{fau}
\end{equation}
When $V(x)$ is periodic, thermodynamic arguments such as those of ref. \cite{LANDAU} allow us to obtain an effective equation for the envelope $\psi$ of the wavefunction $\Psi$, which turns out to be 
\begin{equation}
{\rm i}\frac{\partial\psi}{\partial t}+\frac{1}{2}\nabla^2\psi+g\vert\psi\vert^2\psi=0.
\label{ferrfau}
\end{equation}
In particular, the above equation is potential--free: $V^{\rm eff}=0$.  However, the approach of \cite{LANDAU} breaks down when the potential $V$ in (\ref{fau}) fails to be periodic; such is the case when $V(x)$ is quasiperiodic, for example, within a quasicrystal. The question arises, what is the effective equation for the long--range amplitudes $\psi$ corresponding to $\Psi$ when $V(x)$ in (\ref{fau}) is quasiperiodic? 

An example of a quasiperiodic potential in 1 dimension is \cite{BESICOVITCH}
\begin{equation}
V_1(x)=\sin x + \sin\omega x
\label{tta}
\end{equation}
where $\omega$ is {\it irrational}. (Actually, the key point to quasiperiodicity is the irrationality of the {\it ratio}\/ of frequencies, in our case $\omega/1=\omega$, but we can always normalise one frequency to unity). One can regard quasiperiodicity as the result of projecting the 2--dimensional periodic potential
\begin{equation}
V_2(x,y)=\sin x+\sin y
\label{dostta}
\end{equation}
onto the 1--dimensional subspace defined by $y=\omega x$:
\begin{equation}
V_1(x)=V_2(x,\omega x).
\label{proy}
\end{equation}
In general, any periodic potential in $2n$ dimensions can be written as a sum (possibly infinite) of terms such as those present in (\ref{dostta}), plus possibly cosine functions.  For simplicity we will restrict our attention to quasiperiodic potentials $V_n(x)$ in $n$ space dimensions, of the type
\begin{equation}
V_n(x^1, \ldots, x^n)=\sum_{i=1}^n\left(\sin x^i+\sin\omega_i x^i\right), \quad \omega_i\notin\mathbb{Q}.
\label{averg}
\end{equation}
More general quasiperiodic potentials can be treated similarly. The above is the result of projecting a periodic potential $V_{2n}(x,y)$ in $2n$ space dimensions
\begin{equation}
V_{2n}(x^1, \ldots, x^n; y^1,\ldots, y^n)=\sum_{i=1}^{n}\left( \sin x^i+\sin y^i\right)
\label{doppel}
\end{equation}
onto $n$ dimensions, by means of appropriate projection conditions such as
\begin{equation}
y^i=\omega_i x^i, \quad \omega_i\notin\mathbb{Q}, \quad i=1,\ldots, n.
\label{ketion}
\end{equation}

We claim that, under certain circumstances, one can trade an interacting theory ($V\neq 0$) on $n$--dimensional commutative space, for a potential--free theory ($V=0$) on $2n$--dimensional noncommutative space. For an arbitrary potential $V(x)$ such a tradeoff is generally not possible. However there exists one limit of the interacting theory (\ref{fau}) in which this tradeoff becomes possible. Namely, the case when $V(x)$ is {\it quasi}\/periodic and one considers long--wavelength excitations only. We will prove that such is the effective, low--energy limit of the theory corresponding to a quasiperiodic potential $V(x)$ in (\ref{fau}): the nonlinear Schroedinger equation (\ref{ncmmnlsq}) on noncommutative spacetime, with twice as many dimensions. Roughly speaking, quasiperiodicity of $V(x)$ can be traded for space noncommutativity of the sort (\ref{kett}), while the {\it microscopic} wavefunction $\Psi$ is replaced with its {\it effective, long--range} envelope $\psi$. In what follows we prove the previous statement. The proof is carried out in two steps. We first determine the required noncommutativity tensor $\theta^{ij}$; this is done below. Next, in section \ref{beweis}, we prove that the effective potential does vanish.

In order to understand why the unprojected $2n$--dimensional theory must be noncommutative we observe that the irrationality of the ratios $\omega_i/1=\omega_i$ ensures the quasiperiodicity of the potential $V_n(x^i)$ along each dimension $x^i$, for all $i=1,\ldots, n$. Quasiperiodicity is the hallmark of noncommutativity \cite{YPMA}. 
However there is nothing in the $n$--dimensional, quasiperiodic theory that reminds one of noncommutativity: the $x^i$ certainly commute among themselves.
Only through the introduction of the new coordinate $y^i$ in eqns. (\ref{doppel}), (\ref{ketion}) do we arrive at a noncommutative space \cite{YPMA}: 
\begin{equation}
[X^I,X^J]={\rm i}L^2\theta^{IJ},\qquad I,J=1,\ldots, N
\label{poeh}
\end{equation}
where $X^{I}$ collectively denotes $x^{i}, y^{i}$. This implies that the appropriate multiplication operation in the unprojected, $2n$--dimensional theory is the $\star$--product (\ref{stern})---which noncommutativity tensor $\theta^{IJ}$ enters the latter has to be determined. We first observe that the indices $I,J$ must run over the range $1,2,\ldots, N=2n$. In fact $\theta^{IJ}=0$ except when $I$ denotes $x^i$ and $J$ denotes the corresponding $y^i$ for the same value of $i$, as in (\ref{doppel}). Thus the only nonvanishing entries of the noncommutativity tensor can be labelled $\theta^{x^iy^i}$, for $i=1,\ldots, n$. Now $\theta^{x^iy^i}$ must depend on the frequencies $\omega_i$ and $1$ (whose ratio $\omega_i/1=\omega_i$ is irrational in the quasiperiodic case), because these are the only data at hand, while it cannot depend on any of the other  $\omega_l$ when $l\neq i$.  We may therefore assume that $\theta^{x^iy^j}=\theta^{x^iy^i}(\omega_i)$. The argument of the latter function is however dimensionless, because $\omega_i$ is in fact the ratio of frequencies $\omega_i/1$. Moreover, this function must vanish whenever $\omega_i$ is rational. It must  also be antisymmetric in $x^i, y^i$. Let us see how a tensor meeting all these requirements can be constructed.

We are looking for a function $\theta^{x^iy^i}(\omega_i)$ that will vanish whenever $\omega_i$ is rational. Since the rationals are dense within the reals, any continuous function vanishing whenever $\omega_i\in\mathbb{Q}$ necessarily vanishes identically on $\mathbb{R}$. This implies that one must necessarily renounce continuity of $\theta^{x_iy_i}$, at least on {\it all}\/ of $\mathbb{R}$, if the noncommutativity tensor is not to be identically zero. We recall that the Dirichlet function $D:\mathbb{R}\rightarrow\mathbb{R}$
\begin{equation}
D(x):=
\left\{\begin{array}{ll}
0,\quad \mbox{$x\in\mathbb{Q}$}\\
1,\quad\mbox{$x\notin\mathbb{Q}$,} 
\end{array}\right.
\label{acaedms}
\end{equation}
is discontinuous everywhere on $\mathbb{R}$. However it can be made continuous at a suitable set of isolated points within the real line, by appropriately multiplying it with some continuous function. For example, the function $xD(x)$ is continuous at $x=0$ and discontinuous everywhere else. If $p(x)$ is a polynomial, then $p(x)D(x)$ is continuous at the zeroes of $p(x)$. As an additional example, the function $\sin(\pi x)D(x)$ is continous at every $x\in\mathbb{Z}$ and discontinuous everywhere else. There is however no way to make $f(x)D(x)$ continuous and nonvanishing in an open, connected set, for any function $f(x)$. These arguments lead one to define
\begin{equation}
\theta^{x^iy^i}(\omega_i):=\omega_iD(\omega_i),
\label{comtessa}
\end{equation}
while antisymmetry can be achieved if we set, by definition, 
\begin{equation}
\theta^{y^ix^i}(\omega_i):=-\theta^{x^iy^i}(\omega_i).
\label{madded}
\end{equation}
The choice of the function $f(\omega_i)=\omega_i$ multiplying the Dirichlet function in (\ref{comtessa}) is such that it ensures continuity at the origin $\omega_i=0$. Admittedly, there are an infinite number of functions that one can pick to multiply the Dirichlet functions on the right--hand side while satisfying our requirements. However, continuity at the origin seems to be natural to require, since we are deforming a periodic theory at $\omega_i=0$ into a quasiperiodic one when $\omega_i\notin\mathbb{Q}$. Our function $f(\omega_i)=\omega_i$ is the simplest choice.

Having specified the necessary $\star$--product to work with, we can now return to eqn. (\ref{doppel}) and write it as one should on noncommutative space, namely
\begin{equation}
V_{2n}(x^1, \ldots, x^n; y^1,\ldots, y^n)=\sum_{i=1}^{n}\left( \sin_{\star} x^i+\sin_{\star} y^i\right),
\label{stardoppel}
\end{equation}
where the function $\sin_{\star}(z)$ is defined by its power series expansion in $\star$--products,
\begin{equation}
\sin_{\star}(z):=\sum_{j=1}^{\infty}\frac{(-1)^{j-1}}{(2j-1)!}z\star z\star {2j-1\atop\cdots} \star z.
\label{estar}
\end{equation}
In other words, all multiplications within the potential are to be performed with the $\star$--product. Fortunately, by the discussion leading up to eqn. (\ref{comtessa}), any potential function whose variables are {\it separated}\/ (such as our case (\ref{stardoppel})) reduces to its commutative expression (\ref{doppel}), and we may continue to use the latter without worrying about (\ref{stardoppel}).  Of course, kinetic terms corresponding to the new coordinates $y^{i}$ arising in the potential $V_{2n}(x,y)$ must also be included in the noncommutative Lagrangian. However, by eqn. (\ref{menta}) (see also (\ref{ncakknonmm}), (\ref{ncmmnlsq})), kinetic terms in the $y^{i}$ are not affected by the noncommutativity. Moreover, kinetic terms will not mix the $x^{i}$ with the $y^{i}$, and separated variables will continue to be separate. Hence the extra kinetic terms introduced by the passage to noncommutative space can be integrated out and disposed of in the normalisation integral of the effective wavefunction $\psi$. It must be borne in mind, however, that this neat separation of variables need not be true for arbitrary potentials.

\section{Proof that the effective potential vanishes}\label{beweis}

As already remarked, a key point is the disappearance of the interaction potential in the effective theory for the envelope wavefunction $\psi$:  $V^{\rm eff}=0$  \cite{LANDAU}. In other words, the envelope wavefunction $\psi$ satisfies a potential--free Schroedinger equation. This can be represented diagrammatically as the passage from the microscopic theory described by the wavefunction $\Psi$ to the effective theory described by its envelope $\psi$:
$$
\;{\rm microscopic}\qquad{\rm effective}
$$
\begin{equation}
V_{k}\neq 0\quad\longrightarrow\quad V^{\rm eff}_{k}=0.
\label{hrzt}
\end{equation}
The arrow stands for the operation of taking the effective limit of the theory in any number $k$ of commutative dimensions. Alternatively we can move vertically along the diagram
$$
{\rm noncommutative}\;\; V_{2n}\neq 0
$$
\begin{equation}
\qquad\qquad\qquad\qquad{\downarrow}
\label{vvrt}
\end{equation}
$$
{\rm commutative}\qquad\;\; V_{n}\neq 0,
$$
where the arrow stands for the projection from $2n$ noncommutative dimensions onto $n$ commutative dimensions. Finally we are interested in a diagram such as
$$
\qquad\qquad\qquad\qquad\quad{\rm microscopic}\quad{\rm effective}
$$
$$
{\rm noncommutative}\quad\quad V_{2n}\neq 0\longrightarrow V^{\rm eff}_{2n}=0
$$
\begin{equation}
\qquad\qquad\qquad\qquad\qquad{\downarrow}\qquad\quad\qquad{\downarrow}
\label{ddgg}
\end{equation}
$$
{\rm commutative}\qquad\qquad V_{n}\neq 0\longrightarrow V^{\rm eff}_{n}=0.
$$
By consistency, the diagram must produce the same result regardless of the order followed, {\it i.e.}, whether one moves first vertically and then horizontally, or viceversa. Whenever a diagram such as (\ref{ddgg}) exists, the resulting effective theory for the envelope wavefunction $\psi$ is potential--free and well defined both in $2n$ noncommutative dimensions and in $n$ commutative dimensions. We will prove that diagram (\ref{ddgg}) exists. Indeed the vertical legs are given by the projection (\ref{ketion}); our task is to construct the missing horizontal legs. 

Consider the top horizontal arrow in (\ref{ddgg})
\begin{equation}
V_{2n}\neq 0\longrightarrow V^{\rm eff}_{2n}=0,
\label{nancyzz}
\end{equation}
The approach of ref. \cite{LANDAU} for constructing the effective theory is inapplicable to (\ref{nancyzz}), since it was based on a {\it commutative}\/ space. Also, the bottom horizontal arrow in 
(\ref{ddgg}),
\begin{equation}
V_{n}\neq 0\longrightarrow V^{\rm eff}_{n}=0,
\label{nancy}
\end{equation}
as constructed in ref. \cite{LANDAU}, relied critically on the {\it periodicity}\/ properties of the potential; therefore it also fails when applied to (\ref{nancy}). We are looking for a definition of the effective theory for the envelope wavefunction $\psi$ that is potential--free and well defined both in $2n$ noncommutative dimensions, where the potential acting on $\Psi$ is periodic, and in $n$ commutative dimensions, where the potential acting on $\Psi$ is quasiperiodic.

Any $\omega\in\mathbb{R}$ can be arbitrarily approximated by a sequence $\{w_l\}$ of rationals,
\begin{equation}
\lim_{l\to\infty}w_l=\omega, \qquad w_l\in\mathbb{Q}\quad\forall l.
\label{sqenc}
\end{equation}
In particular this implies that any quasiperiodic potential $V_n(x)$ can get as close to being periodic as one wishes (this is in fact one possible definition of {\it quasiperiodicity}\/, if one replaces the term {\it potential function}\/ with an arbitrary {\it function}\/  \cite{BESICOVITCH}). Let now $\{W_{l,n}(x)\}$ denote a sequence of $n$--dimensional periodic potentials corresponding to the rational sequence $\{w_l\}$, and such that
\begin{equation}
\lim_{l\to\infty}W_{l,n}(x)=V_{n}(x).
\label{zassow}
\end{equation}
Note that we write $W_{l,n}(x)$ rather than $W_{l,n}(x,y)$ because these potentials, although periodic, depend only on $n$ commutative space coordinates $x$. Let now $W^{\rm eff}_{l,n}(x)$ denote the effective potential corresponding to $W_{l,n}(x)$, obtained as per ref. \cite{LANDAU}. It follows that eqn. (\ref{nancy}) can be approximated as
\begin{equation}
W_{l,n}(x)\neq 0\longrightarrow W^{\rm eff}_{l,n}(x)=0.
\label{uvedoble}
\end{equation}
Therefore we can approximate the right--hand side of (\ref{nancy}) as
\begin{equation}
\lim_{l\to\infty}W^{\rm eff}_{l,n}(x)=V^{\rm eff}_{n}(x)
\label{chance}
\end{equation}
but, since the sequence $\{W^{\rm eff}_{l,n}(x)\}$ is identically zero, its limit must also vanish:
\begin{equation}
V_{n}^{\rm eff}(x)=0.
\label{kakao}
\end{equation}
We have thus proved eqn. (\ref{nancy}). It still remains to prove that (\ref{nancyzz}) is true. This follows from the fact that the projection conditions (\ref{ketion}) can only lead to a vanishing $V_n^{\rm eff}$ downstairs if $V_{2n}^{\rm eff}$ upstairs is itself zero. We conclude that the diagram (\ref{ddgg}) exists and commutes. 

Proving the existence of the horizontal arrow (\ref{nancy}) was a simple exercise in the properties of rational numbers. One could ask, why complicate matters by lifting the whole construction to $2n$ noncommutative dimensions? The answer is simple: noncommutativity is imposed on us by quasiperiodicity \cite{YPMA}. Thus our construction can be regarded as a {\it definition}\/ of the effective theory on noncommutative space. By the UV/IR mixing of noncommutative theories \cite{SZABO}, it is not {\it a priori}\/ evident how an effective theory is to be defined on noncommutative space. Our prescription thus amounts to the following: whenever noncommutativity has its origin in the quasiperiodicity of the potential, the effective theory on noncommutative space is the lift of the corresponding effective theory on commutative space. If the latter has a length resolution of value $L$, then $L^2$ must appear on the right--hand side of the commutation relations (\ref{poeh}) on noncommutative space.

\section{Discussion}\label{konk}

We have proved that  low--energy, nonlinear Schroedinger equation (\ref{fau}) for a wavefunction $\Psi$ in the presence of a quasiperiodic potential gives rise to the nonlinear Schroedinger equation (\ref{ncmmnlsq}), on noncommutative spacetime, for the envelope wavefunction $\psi$ of $\Psi$, the latter equation carrying no potential. Thus the effect of a quasiperiodic potential on commutative space can be mimicked by the passage to a potential--free theory on noncommutative space in twice as many dimensions.

The simplest example of a noncommutative space is given by eqn. (\ref{kett}), where the coordinates satisfy a Heisenberg--like algebra instead of being commutative. It can be proved \cite{CONNES, SZABO} that the commutative space $\mathbb{R}^n$ is rendered noncommutative under the replacement of the usual, commutative, pointwise product of functions $f(x)\cdot g(x)$ with the noncommutative $\star$--product of eqn. (\ref{stern}). This replacement causes the nonlinear Schroedinger equation (\ref{nlsq}) to become its noncommutative analogue (\ref{ncmmnlsq}). The latter  differs from the former by the $\star$--products in the cubic terms only, quadratic terms in the wavefunction being insensitive to the $\star$--product thanks to eqn. (\ref{menta}).

In order to specify a $\star$--product on $\mathbb{R}^n$ it is necessary and sufficient to determine a constant antisymmetric tensor $\theta^{ij}$ as in (\ref{stern}). On the other hand, a quasiperiodic potential in $n$ dimensions is specified by a set of $n$ irrational frequencies $\omega_i\in\mathbb{R}$. Once the latter are known, the tensor $\theta^{ij}$ is determined, as a function of the frequencies defining the potential, by eqn. (\ref{comtessa}).  Although we have concentrated our attention on quasiperiodic potentials of the type (\ref{averg}), our technique easily extends to more general quasiperiodic potentials. For example, if the quasiperiodic potential reads
\begin{equation}
V_1(x)=\sin x+\sin\omega x+\sin\xi x, \qquad \omega, \xi\notin\mathbb{Q},
\label{kompliziert}
\end{equation}
then we consider
\begin{equation}
V_3(x,y,z)=\sin x+\sin y+\sin z, \qquad y=\omega x,\; z=\xi x.
\label{den}
\end{equation}
This leads to nonzero commutators $[x,y]$ and $[x,z]$. If, moreover, the irrationals $\omega$ and $\xi$ are incommensurable, then $[y,z]$ will also be nonzero. Therefore, in general, the noncommutative theory has more than just {\it twice}\/ as many dimensions as the original theory.

After completion of this manuscript we became aware of ref. \cite{HORVATHY}, where issues closely related with ours are discussed from an interesting alternative perspective.

\vskip1cm
\noindent {\bf Acknowledgements}  J.M.I.  thanks Max--Planck--Institut f\"ur Gravitationsphysik, Albert--Einstein--Institut (Golm, Germany) for hospitality during the final stages of this article. This work has been supported by Generalitat Valenciana (Spain).

\end{document}